\documentclass[preprint,12pt]{elsarticle}

\usepackage{amssymb}
\usepackage{adjustbox} %Added by me
\usepackage{amsmath,latexsym,amssymb,amsfonts}
\usepackage{float} \usepackage{graphicx}% Include figure files
\usepackage{graphics}
\usepackage{caption}
\usepackage{subfigure}
\usepackage{bm}
\usepackage{color}

\usepackage[dvips]{psfrag}
\usepackage{ulem}
\usepackage{epsfig}
\usepackage{cases}
\usepackage{tensor}
\usepackage{multirow,array}
\usepackage{booktabs} %For prettier tables

\journal{Physics of Dark Universe}

\begin{document}

\begin{frontmatter}



\title{The cosmological evolution condition of the Planck constant in the varying speed of light models through adiabatic expansion }


\author{Seokcheon Lee}

\affiliation{organization={Department of Physics, Institute of Basic Science},%Department and Organization
            addressline={Sungkyunkwan University}, 
            city={Suwon},
            postcode={16419}, 
            %state={},
            country={Korea}}

\begin{abstract}
There have been various varying speed of light (VSL) models with one free parameter, $b$, to characterize the time variation of the speed of light as a function of a scale factor, $c = c_0a^{b/4}$, based on the expanding universe. One needs to induce cosmological evolutions of other physical constants and quantities having different powers of scale factor as a function of $b$ to satisfy all known local physics laws, including special relativity, thermodynamics, and electromagnetic force. These models should be based on the Friedmann-Lema\^{i}tre-Robertson-Walker metric satisfying the isotropic and homogeneous three-space known as the cosmological principle. Adiabaticity is a necessary condition to keep homogeneity and isotropy because a net energy flux would falsify the isotropy if there is a preferential energy flow direction. It also might forge homogeneity if the outward (inward) flow is isotropic. Thus, any VSL model based on the expanding universe should preserve an adiabatic expansion condition to be a viable model. We show that this condition specifies the cosmological evolution of the Planck constant as $\hbar = \hbar_0 a^{-b/4}$.  
\end{abstract}


%\item Cosmological time-varying speed of light model
%\item Adiabatic expansion of the Universe


\end{frontmatter}


\section{Introduction}
\label{sec:intro}

The spatial distribution of galactic clusters shows apparent isotropy and statistical homogeneity on scales exceeding $250$ million light years. Astronomers have made estimations regarding the cosmic scale at which the transition from a lumpy, inhomogeneous Universe to a smoother, statistically homogeneous Universe takes place, utilizing cosmological observations. Some of these estimations can be found in \cite{2010MNRAS.405.2009Y,2022JCAP...10..088A}. For further details, please refer to the references therein as well. This observed characteristic can be effectively described by the Friedmann-Lema\^{i}tre-Robertson-Walker (FLRW) metric, which provides a suitable framework for understanding the underlying spacetime structure. Adiabatic expansion, in the context of cosmology, denotes a process where there is no exchange of heat with the surroundings, and any change in the system's internal energy is solely determined by work. Adiabaticity plays a crucial role in maintaining the homogeneity and isotropy mandated by the cosmological principle (CP), which asserts the uniformity of the universe on large scales.

%The distribution of galactic clusters in space appears to be isotropic and statistically homogeneous on scales larger than 250 million light years. The Robertson-Walker (RW) metric well describes this spacetime. Adiabatic expansion refers to one in which no heat is transferred into or out of a system, and the change in internal energy is determined by only work. Adiabaticity is a necessary condition to keep the homogeneity and the isotropy required by the cosmological principle (CP). 

The first law of thermodynamics, known as the law of conservation of energy, can be expressed as $dQ = dE + P dV$, where $dQ$ represents the heat flow into or out of a given volume. In the case of a perfectly homogeneous and isotropic Universe, the condition $dQ = 0$ holds true for any volume, indicating the absence of bulk heat flow (\textit{i.e.},  adiabaticity). In such a scenario, a homogeneous and isotropic expansion of the Universe does not contribute to an increase in the overall entropy of the Universe.

%The first law of thermodynamics (law of conservation of energy),  $dQ = dE + P dV$, where $dQ$ is the heat flow into or out of a volume. If the Universe is perfectly homogeneous and isotropic, then for any volume $dQ = 0$, that is, there is no bulk flow of heat (\textit{i.e.},  adiabatic). A homogeneous, isotropic expansion of the Universe does not increase the entropy of the Universe. 

It follows that the time evolution of the cosmic microwave background (CMB) temperature for an adiabatic expansion predicts the linear increase with redshift, $T(z) = T_0 (1+z)$, where $T_0$ is the present value of the CMB temperature. There are several methods to measure this theoretical prediction at different redshifts. The CMB photons interact with the hot intracluster medium (ICM), producing the thermal Sunyaev-Zeldovich (tSZ) effect \cite{1970Ap&SS...7....3S,1980ARA&A..18..537S}. This inverse Compton scattering of CMB photons by electrons provides a direct measurement of the CMB temperature at redshift $0 \leq  z \leq 1$. One can also use the fact that CMB radiation excites the rotational lines of molecules and/or atoms in quasar (QSO) absorption line systems at redshifts $1.8 \leq z \leq 3.3$, but this is model-dependent \cite{1968ApJ...152..701B}. One can also use a massive starburst galaxy to measure $T$ at higher redshifts \cite{2022Natur.602...58R}.  Any deviation from the adiabatic expansion of the Universe can be expressed by the modification of the redshifting of the CMB temperature as $T(z) = T_0 (1+z)^{1-\beta}$ \cite{2000MNRAS.312..747L}.  

There have been several data from different missions to constrain the value of $\beta$ shown in Table~\ref{Tab:1}.  The South Pole Telescope (SPT) obtains measurements of the spectrum of the Sunyaev-Zeldovich effect (SZe) using the ratio of the SZ signal at $95$ and $150$ GHz to constrain deviations from the expected adiabatic evolution of CMB temperature \cite{2014MNRAS.440.2610S}.  They apply the method to a sample of $158$ SPT-selected clusters covering a redshift range of $0.05 < z < 1.35$. There is another SZe spectrum obtained from Planck temperature maps at various frequencies ranging from 70 to 353 GHz for a subset of 104 clusters from the Planck SZ cluster catalog. Using a Monte-Carlo Markov Chain approach and examining the SZ intensity change at different frequencies, they derived individual measurements of CMB temperature for each cluster in the sample \cite{2015JCAP...09..011L}. By analyzing data from $370$ clusters obtained from the largest SZ-selected cluster sample to date, collected by the Atacama Cosmology Telescope (ACT), they derive new constraints on the deviation of CMB temperature evolution from the standard model (SM \textit{i.e.}, $c =$ const) \cite{2021ApJ...922..136L}. All these results are consistent with $\beta = 0$, which corresponds to the adiabatic expansion. 

 \begin{table*}[h!]
 	\centering
%\begin{adjustbox}{width=1\textwidth}
		%\small
\begin{tabular}{ |c|c|c|c| } 
 \hline
 & SPT &  Planck DR$1$ & ACT \\ 
\hline
$\beta$ & $0.017_{+0.030}^{-0.028}$ &  $0.012 \pm 0.016$ 
& $0.017_{-0.032}^{+0.029}$ \\ 
\hline
ref & \cite{2014MNRAS.440.2610S}     & \cite{2015JCAP...09..011L}  & \cite{2021ApJ...922..136L}  \\ 
\hline
\end{tabular}
%\end{adjustbox}
\caption{These are the values of $\beta$ obtained from various missions.}
\label{Tab:1}
 \end{table*} 

There have been various VSL models \cite{1999PhRvD..59d3515B,1988MPLA....3.1527P,1995ApSS.226..273P,1999PhRvD..59d3516A,2007CaJPh..85.1395S,2006CaJPh..84..933S,2008arXiv0803.1362P,2021JCAP...08..054L,2023FoPh...53...40L}. We can obtain an extended theory satisfying both the Lorentz invariance (LI) and the law of energy conservation even when the speed of light varies as a function of cosmic time $c(a)$. One needs to obtain the cosmological evolutions of other physical constants to satisfy LI, electromagnetism, and thermodynamics. The so-called minimally extended varying speed of light (meVSL) model satisfies these local physics laws  \cite{2021JCAP...08..054L,2023FoPh...53...40L}. We compare the cosmological evolutions on physical constants and quantities between different VSL models in Table~\ref{tab:tabVSL}.  We especially focus on the condition of the cosmological evolution of the Planck constant to preserve the CP to clarify the viability of VSL models. %from Landau:2000mx p.2

 \begin{table*}[h!]
 	\centering
		\begin{adjustbox}{width=1\textwidth}
		%\small
		\begin{tabular}{|c|c|c|c|c|c|c|c|c|c|c|} 
			\hline
			 $c$ & $G$ & $\hbar$ &$\lambda$ & $\nu$  &  $m$ & $k_{\textrm{B}}$ & $T$ & $e$  & $\alpha$   &reference \\ \hline
		 $c_0 a^{-1/2}$  & $G_0 a^{-1}$ & $\hbar_0 a^{3/2}$ & $\lambda_0 a$ & $\nu_0 a^{-3/2}$ & $m_0 a$ & NC & NC & $e_0 a^{1/2}$ & const & \cite{1988MPLA....3.1527P,1995ApSS.226..273P,2008arXiv0803.1362P} \\
			 $c_0 a^n$ & const & $\hbar_0 a^{n}$ &const & $\nu_0 a^n$ & const & const & $T_0 a^{2n}$ & const & $\alpha_0 a^{-2n}$  & \cite{1999PhRvD..59d3515B,1999PhRvD..59d3516A} \\
			 $c_0 a^{-1/4}$ & const & NC & $\lambda_0 a$ & $\nu_0 a^{-5/4}$ & $m_0 a^{1/2}$ & const & $T_0 a^{-5/4}$ & NC & NC  & \cite{2007CaJPh..85.1395S,2006CaJPh..84..933S} \\
			  $c_0 a^{b/4}$ & $G_0 a^{b}$ & $\hbar_0 a^{-b/4}$ & $\lambda_0 a$ & $\nu_0 a^{-1+b/4}$ & $m_0 a^{-b/2}$ & const & $T_0 a^{-1}$ & $e_0 a^{-b/4}$ & $\alpha_0 a^{-b/4}$  & \cite{2021JCAP...08..054L,2023FoPh...53...40L} \\
			\hline
		\end{tabular}
		\end{adjustbox}
		\caption{Cosmological evolutions on physical quantities and constants of various VSL models. NC means not considered. }
		\label{tab:tabVSL}
 \end{table*} 

Any viable VSL model should also guarantee adiabatic expansion to preserve the CP in addition to satisfying all known local physics laws. Both theories and observations minimally require this condition. Thus, we should exclude any cosmological evolution relations of physical quantities and constants of VSL models which violate the adiabatic expansion condition. 

Testing the stability of fundamental couplings in nature offers valuable insights into new physics. Detecting variations in these couplings would be groundbreaking, while even improved null results place competitive constraints on various cosmological and particle physics theories \cite{2011LRR....14....2U,2017RPPh...80l6902M,2022arXiv221013187B,2023MNRAS.521..850L,2023arXiv230204565S}.  Thus, there are plans and forecasts for studies using advanced facilities like ALMA \cite{2013arXiv1309.3519F,2014arXiv1406.4650T}, ESPRESSO \cite{2014MmSAI..85..149P}, ELT \cite{2019arXiv190202785L},  Euclid \cite{2013LRR....16....6A},  CORE \cite{2018JCAP...04..015D}, etc.

 In Sec.\ref{sec:adiRW}, we review how to derive the cosmological evolution of the CMB temperature from the adiabatic expansion. We investigate whether the adiabatic expansion condition is obtained in various VSL models or not in section~\ref{sec:adimeVSL}. If this condition is not satisfied, it is not a viable model, even if it can satisfy all known local physics laws due to the violation of the CP. We conclude in Sec. \ref{sec:Conc}. 

\section{Review of Adiabatic expansion}
\label{sec:adiRW}

The first law of thermodynamics is a statement of conservation of energy. In the momentarily comoving reference frame, the fluid element can exchange energy with its surroundings in two ways: by heat conduction (absorbing heat) and by work (doing work). Let $d Q$ be heat energy gained,  $E$ is the total energy of the element, and $ P dV$ is energy lost, then one can write 
\begin{equation}
dQ = dE + P dV = d \left( \varepsilon V \right) + P dV \label{2ndlaw} \,.
\end{equation}
Processes for which $dQ = 0$ are known as adiabatic processes. The adiabatic expansion of the Universe does not increase the entropy of the Universe.  Because the photon is a dominant component to contribute entropy,  we consider 
\begin{align}
\varepsilon_{\gamma} = \frac{\pi^2}{15} \frac{\left( k_{\textrm{B}} T_{\gamma} \right)^4}{\left( \hbar c \right)^3} \equiv \sigma_{\gamma} T^4 \quad , \quad \sigma_{\gamma} = \frac{\pi^2}{15} \frac{k_{\textrm{B}}^4}{\left( \hbar c \right)^3}
\quad , \quad p_{\gamma} = \frac{1}{3} \varepsilon_{\gamma} \label{pgamma} \,,
\end{align}
where $\sigma_{\gamma}$ is the so-called black-body constant, related to the Stefan-Boltzmann constant, $\sigma_{SB}$, as $\sigma_{SB} = \sigma_{\gamma} c/4$.   One can use Eq.\eqref{pgamma} into Eq.~\eqref{2ndlaw} to obtain 
\begin{align}
d Q = 4 \sigma_{\gamma} T_{\gamma}^4 V_0 a^3 \left[ d \ln T_{\gamma} + d \ln a \right]  \label{ConS} \,,
\end{align}
where $a$ is a scale factor.  If we adopt the adiabatic expansion condition $dQ = 0$, then we obtain the time evolution of the CMB temperature
\begin{align}
T_{\gamma} = T_{\gamma 0} a^{-1} = T_{\gamma 0} (1+z) \label{Tzadi} \,,
\end{align}
where we use $a_0 =1$ and $z$ is the cosmlogical redshift.

\section{Adiabatic expansion in VSL models}
\label{sec:adimeVSL}

We can repeat the consideration in the previous section \ref{sec:adiRW} by including the time evolution of cosmological constants. In these cases, the black-body constant can depend on cosmic time to obtain Eq~\eqref{ConS} for the adiabatic expansion as 
\begin{align}
4 \sigma_{\gamma} T_{\gamma}^4 V_0 a^3 \left[ d \ln T_{\gamma} + d \ln a + \frac{1}{4} d \ln \sigma_{\gamma} \right] = 0 \label{VSLS} \,.
\end{align}
One can define the different time-dependent black-body constants for various VSL models. We investigate these in this section.

\subsection{mVSL}
\label{subsec:mVSL}
A VSL model which proposed the change of the speed of light only without allowing the variations of other physical constants is called the minimal VSL (mVSL) \cite{1993IJMPD...2..351M,1999PhLB..460..263C,1999CQGra..16.1435B}. By assuming $c = c_0 a^{b/4}$,  the black-body constant becomes
\begin{equation}
\sigma_{\gamma} = \frac{\pi^2}{15} \frac{k_{\textrm{B} 0}^4}{\left( \hbar_{0} c_{0} \right)^3} a^{-\frac{3b}{4}} = \sigma_{\gamma 0}  a^{-\frac{3b}{4}} \label{sigmamVSL} \,,
\end{equation}
where the Boltzmann constant does not vary. One can obtain the cosmological evolution of CMB temperature by plug Eq.~\eqref{sigmamVSL} into Eq.~\eqref{VSLS} 
\begin{align}
T_{\gamma} = T_{\gamma 0} a^{-1 +\frac{3b}{16}} \equiv T_{\gamma 0} a^{-1 +\beta} \label{TCMBmVSL} \,.
\end{align}
Thus, there exists the modification of the redshifting of the CMB temperature in this model. 

\subsection{meVSL}
\label{subsec:meVSL}
Cosmological evolutions of physical constants and quantities of the meVSL model are summarized in Table~\ref{tab:tabVSL}. From these, we can establish consistent all known local physics laws, including special relativity, thermodynamics, and electromagnetism.  In this model, the black-body constant is the same as the SM one
\begin{align}
\sigma_{\gamma} = \frac{\pi^2}{15} \frac{k_{\textrm{B} 0}^4}{\left( \hbar_{0} c_{0} \right)^3} = \sigma_{\gamma 0} \label{sigmameVSL} \,.
\end{align}
Thus, the cosmological redshift of the CMB temperature is also consistent with that of SM in the meVSL model
\begin{align}
T_{\gamma} = T_{\gamma 0} a^{-1} \label{TCMBmeVSL} \,.
\end{align}

\subsection{Other VSL}
\label{subsec:OVSL}
One can also obtain the different time-dependent black-body constants for other VSL models as
\begin{align}
 \sigma_{\gamma} = \frac{\pi^2}{15} \frac{k_{\textrm{B} 0}^4}{\left( \tilde{\hbar}_0 \tilde{c}_0 \right)^3} a^{-\frac{3b_2}{4}} \equiv \sigma_{\gamma 0}a^{-\frac{3b_2}{4}} \quad \textrm{if} \quad \tilde{\hbar} = \tilde{\hbar}_0 a^{\frac{b_1}{4}} \,, \, b_1 \neq b \label{sigmaOVSL} \,,
\end{align}
where $b_2 = b + b_1$ with $b_1 \neq -b$. It causes the deviation of CMB temperature from the linear increase as
\begin{align}
T_{\gamma} = T_{\gamma 0} a^{-1 +\frac{3b_2}{16}} \equiv T_{\gamma 0} a^{-1 +\beta} \label{TOVSL} \,.
\end{align}
Thus, if one investigates the VSL model without considering the accompanying variation of the Planck constant, one obtains the CMB temperature's different time evolution for an adiabatic expansion compared to that of the SM.  The current observations show that $\beta = 0$, and thus one should exclude any VSL model except for $\tilde{\hbar} = \tilde{\hbar}_0 a^{-\frac{b}{4}}$ for $\tilde{c} = \tilde{c}_0 a^{\frac{b}{4}}$ model.  This condition can be extended to encompass any model where the speed of light evolves smoothly with the scale factor, denoted as $\tilde{c} = \tilde{c}_0 f(a)$, by imposing $\tilde{\hbar} = \tilde{\hbar}_0 f^{-1}(a)$, with $f^{-1}(a)$ representing the inverse function of $f(a)$.

\section{Conclusion}
\label{sec:Conc}

In the context of the expanding Universe, cosmological redshift leads to differences in dimensional quantities, such as the wavelength and temperature of photons, between the observed and emitted states. As a result, considering the constancy or variability of dimensional constants, rather than just dimensionless ones, becomes physically meaningful in the context of cosmic expansion. Thus, one can investigate the viable varying speed of light model based on the RW metric, and it should satisfy the isotropy and homogeneity of the three-space, called the cosmological principle. Adiabaticity is a necessary condition to maintain the cosmological principle. From this, we should specify the cosmological evolution of the Planck constant for the given form of the speed of light.  The analysis relies on the assumption of the cosmological principle, which posits the statistical homogeneity and isotropy of the Universe. Recent observations have suggested potential deviations from the expectation of cosmological principle \cite{2023CQGra..40i4001K}. In this case, we need to reconsider the entire paradigm for the Friedmann-Lema\^{i}tree-Robertson-Walker metric with the validity of the adiabatic expansion.

%%%%%%%%%%%%%%%%%%%%%%%%%%%%%%%%%%%%%%%%%%%
\section*{Acknowledgments}
%%%%%%%%%%%%%%%%%%%%%%%%%%%%%%%%%%%%%%%%%%%
SL is supported by Basic Science Research Program through the National Research Foundation of Korea (NRF) funded both by the Ministry of Science, ICT, and Future Planning (Grant No. NRF-2019R1A6A1A10073079) and by the Ministry of Education (Grant No. NRF-RS202300243411). 


%\appendix

%\section{Sample Appendix Section}

 \bibliographystyle{elsarticle-num} 
 \bibliography{cas-refs}





% \end{thebibliography}
\end{document}